\documentclass[twocolumn]{aastex62}
\usepackage{natbib}

\shorttitle{Solar System Hadley Cells}
\shortauthors{Rees \& Garrett}

\begin{document}

\title{Analytical Estimation of the Width of Hadley Cells in the Solar System}

\author{Karlie N. Rees}
\affiliation{Department of Atmospheric Sciences, University of Utah, 135 S 1460 E Rm 819, Salt Lake City, UT 84112, USA}

\author{Timothy J. Garrett}
\affiliation{Department of Atmospheric Sciences, University of Utah, 135 S 1460 E Rm 819, Salt Lake City, UT 84112, USA}

\correspondingauthor{Timothy J. Garrett}
\email{tim.garrett@utah.edu}


\begin{abstract}
The angular width of Earth's Hadley cell has been related to the square root of the product of the tropospheric thickness and the buoyancy frequency, and to the inverse of the square root of the angular velocity and the planetary radius. Here, this formulation is examined for other planetary bodies in the solar system. Generally good consistency is found between predictions and observations for terrestrial planets provided the pressure scale height rather than the tropopause height is assumed to determine the thickness of the tropospheric circulation. For gas giants, the relevant thickness is deeper than the scale height, possibly due to the internal heat produced by Kelvin-Helmholtz contraction. On Earth, latent heat release within deep convection may play a similar role in deepening and widening the Hadley Cell. 
\end{abstract}

\section{Introduction} \label{sec:intro}

The Hadley cell circulation on Earth is its dominant circulation pattern. Approximately axisymmetric, it is bounded by the equator and 30$\degr$ latitude in either hemisphere \citep{Koppen}. In an early effort to derive an analytical expression for the Hadley cell width, \citet{Schneider} considered circulations in an idealized, nearly inviscid atmosphere, assuming that zonal winds are geostrophic, and that they conserve angular momentum with increasing latitude. Provided there is a balance in the cell between latent heating by tropical cumulus and top of the atmosphere infrared cooling, the derived cell width is $y_S \approx (5)^{1/4} (R_L a)^{1/2}$ where $R_L=NH/2\Omega$ is the Rossby radius of deformation for a stratified atmosphere, $a$ is the radius of the Earth, $N$ is the tropospheric Brunt-V\"{a}is\"{a}l\"{a} or buoyancy frequency, $H$ is the depth of the tropical atmosphere, and $\Omega$ is the planetary angular velocity. Through substitution, $y_S\approx (5/4)^{1/4}(aNH/\Omega)^{1/2}$. Converting to an angle through $\phi = y/a$:
\begin{equation}
\phi_S\approx \left(\frac{5}{4}\right)^{1/4}\left(\frac{NH}{\Omega a}\right)^{1/2}
\label{equation:1}
\end{equation}

A similar result was obtained by \citet{Held}, who also assumed that zonal winds conserve angular momentum, but that the cell width is determined by the latitude at which vertical wind shear due to baroclinic instability leads to development of large-scale eddies. Expressing Equation (\ref{equation:1}) in terms of latitude, Held found that $\phi_H \approx (gH\Delta\theta_v/\Omega^2 a^2 \theta_0)^{1/4}$ where $g$ is the gravitational constant, $\Delta\theta_v$ is the vertical change in virtual potential temperature with height and $\theta_0$ is the virtual potential temperature at the surface. Substituting the Brunt-V\"{a}is\"{a}l\"{a} frequency where $N^2=g/H ((\Delta\theta_v)/\theta_0 )$, the expression simplifies to simplifies to:
\begin{equation}
\phi_H\approx \left(\frac{NH}{\Omega a}\right)^{1/2}
\label{equation:2}
\end{equation}

Despite the difference in approaches, Equations (\ref{equation:1}) and (\ref{equation:2}) differ by just 6\%. Both results implicitly express the angular width of the Hadley cell as a ratio of the square root of two velocities: $NH$ is proportional to the square root of the available buoyant potential energy in a stratified atmosphere \citep{Tailleux} and $\Omega a$ is proportional to the square root of the atmosphere's rotational energy. 

Notably, the relevant atmospheric depth was assumed to be that of the troposphere, $H_{trop}$. The reason for this choice is not obvious as it might be argued that the atmospheric density scale height $H_{scale}$  is most directly tied to mass and heat transfer in the general circulation. Here, we explore the general validity of Equation (\ref{equation:2}) for predictions of Hadley cell width by examining a range of planetary bodies in the solar system, paying particular attention to the value of $H$ and exploring the possible dependence of $\phi_{H}$ on internal convective energy sources.

\section{Observed planetary parameters} \label{sec:est}

\begin{deluxetable*}{lcccDDr}[!htb]
\tablenum{1}
\tablecaption{Observed Planetary Parameters}
\tabletypesize{\scriptsize}
\tablehead{ \colhead{Planet} & \colhead{$a$} & \colhead{$N$} & \colhead{$\Omega$} & \multicolumn2c{$H_{scale}$} & \multicolumn2c{$H_{trop}$} & \colhead{$\phi_{obs}$} \\ \colhead{} & \colhead{(km)} & \colhead{($\rm 10^{-2}~s^{-1}$)} & \colhead{($\rm 10^{-5}~rad~s^{-1}$)} & \multicolumn2c{(km)} & \multicolumn2c{(km)} & \colhead{(deg)} 
}
\decimals
\startdata
Venus & 6,050 & 1.05 & 2.31\tablenotemark{a} & 15.9 & 65.0 & 60 \\
Earth & 6,370 & 1.12 & 7.27 & 8.5 & 17.0 & 30 \\
Mars & 3,396 & 0.78 & 7.10 & 11.1 & 45.0 & 40 \\
Jupiter & 71,400 & 1.51 & 17.8\tablenotemark{a} & 27.0 & 124.3 & 18 \\
Saturn & 60,270 & 0.67 & 16.5 & 59.5 & 274.0 & 25 \\
Titan & 2,575 & 0.25 & 0.45 & 40.0 & 50.0 & 170 \\
Uranus & 25,560 & 1.02 & 9.70 & 27.7 & 127.6 & 30 \\
Neptune & 24,760 & 1.33 & 1.62 & 20.0 & 92.1 & 50 \\
\enddata

\tablecomments{$a$: planetary radius, $N$: Brunt-V\"{a}is\"{a}l\"{a} frequency, $\Omega$: planetary rotation rate, $H_{scale}$: pressure scale height, $H_{trop}$: tropopause height, defined for gas giants as the height between 10 and 0.1 bar, $\phi_{obs}$: observed Hadley cell width in degrees. 
\tablenotetext{a}{Adjusted for superrotation}
}
\label{table:1}
\end{deluxetable*}

Table (\ref{table:1}) summarizes planetary parameters relevant to calculation of Equation (\ref{equation:2}) along with the observed width of the Hadley cell. For planetary bodies in the Solar System other than Earth, the latitude $\phi_{obs}$ of the Hadley cell subsidence zones is most often determined through off-equatorial jet analysis supplemented by visual estimates of where cloud banding is related to jet activity \citep{Yamazaki, Showman}. More recently, \citet{Friedson} proposed a Hadley circulation on Saturn with strong subsidence located at $\pm$25$\degr$ latitude using data taken by the \textit{Cassini} spacecraft between 2007 and 2010; a previous estimate used jet analysis to find $\phi_{obs}=30\degr$.  \citet{Bolton} described an Earth-like Hadley circulation on Jupiter with equatorial upwelling and subsequent downwelling occurring between 10 and 20$\degr$ latitude that coincides with a previous jet analysis estimate of $\phi_{obs}=18\degr$. Titan has an unusual  Hadley cell that spans nearly the entire moon during the Saturnian summer and winter seasons, with a semi-permanent polar vortex in the winter hemispheric pole, and that it splits into two as the vortex transitions between poles during the spring and fall over the course of Saturn's 30-year seasonal cycle \citep{Tokano}.

For the value of $\Omega$ in Equation (\ref{equation:2}), all of the planetary giants exhibit some superrotation or subrotation in their atmospheres. On Saturn, Uranus, and Neptune, the subrotating or superrotating jets are thought to be confined to the upper troposphere and only to constitute of order 1\% of the total atmospheric circulation \citep{Showman} and have a negligible influence relative to the planetary body on the total rotation rate of the atmosphere. For Jupiter, the \textit{Juno} spacecraft \citep{Kaspi} revealed surprisingly deep atmospheric jets extending $\sim$3,000\,km into the planet's interior suggesting that a relatively large fraction of the atmosphere is rotating at speeds faster than the planetary rotation rate. Also, Venus has an extremely slow planetary rotation rate of $\rm 3~x~10^{-7}~rad~sec^{-1}$ relative to the atmospheric rotation rate between the surface and the upper-level tropospheric jets \citep{Showman,Ainsworth,Walterscheid}. Thus, for Venus and Jupiter, the atmospheric rotation rate $\Omega$ is assumed to be determined from the strength of the mid-latitude jet divided by the planetary radius. 

The square of the atmospheric buoyancy or Brunt-V\"{a}is\"{a}l\"{a} frequency is approximated from the atmospheric temperature profile through $N^2 = g/T_{e}(\Gamma_{d}-\Gamma)$ where $T_{e}$ is the planetary thermal emission temperature, $\Gamma_{d}=g/c_{p}$ is the dry adiabatic lapse rate specific to the planet where $c_{p}$ is specific heat at constant pressure, and $\Gamma$ is the environmental lapse rate obtained from temperature profiles observed by either radio occultation \citep{Seiff,Rossow} or from probe or lander data \citep{Ainsworth,Eshleman,McKay}. Environmental lapse rates for gas giants are determined from the difference between the average temperature at 1 bar and the average temperature at 0.1 bar \citep{NASA}. 

For all planetary bodies, $H_{scale}$ is the tropospheric pressure scale height \citep{NASA,McKay}. For terrestrial planets, the tropopause height $H_{trop}$ is determined from available soundings \citep{Ainsworth,Eshleman,Brown}, with the exception of Earth, for which an average tropical tropopause height is used. For gas giants, it is related to the scale height from the pressures at 1 mb and 0.1 mb $H_{trop}=H_{scale}ln(1/0.1)$ \citep{RobinsonCatling}.

\section{Comparison of theoretical and observed Hadley cells} \label{sec:comp}

\begin{figure*}[!htb]
\figurenum{1}
\plotone{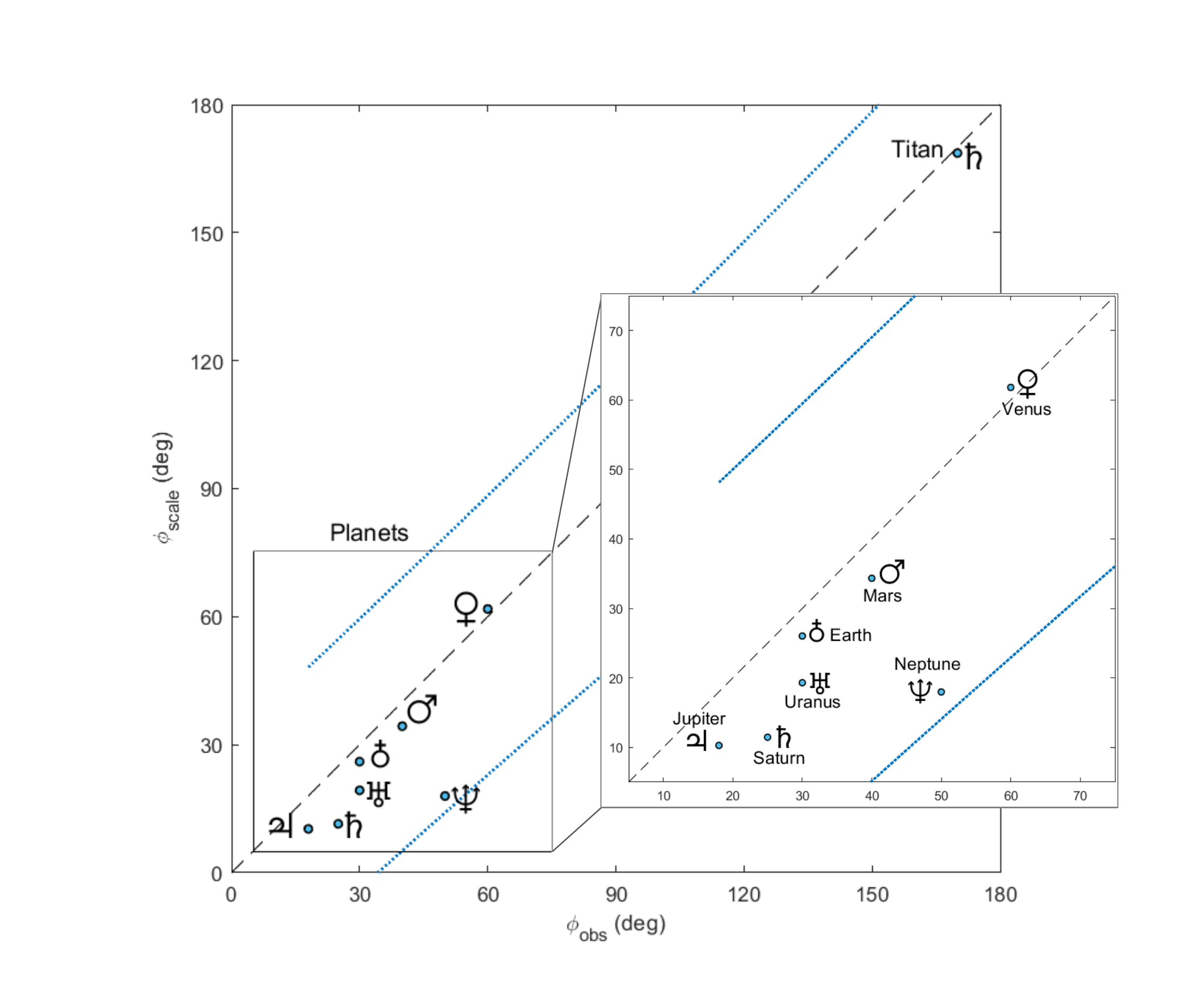}
\caption{Observed values of Hadley cell width versus values calculated with Equation (\ref{equation:2}) using $H_{scale}$ as the value of $H$ and adjusted for superrotation as indicated in Table (\ref{table:1}). Blue dotted lines show 95\% confidence interval bounds for a least-squares best fit of $\phi_{scale}=b\phi_{obs}$ where $b=0.927\pm0.158$. Dashed lines are where $\phi_{obs}=\phi_{scale}$.}
\label{fig:fig1}
\end{figure*}

Figure (\ref{fig:fig1}) shows a comparison of observed values of Hadley cell width with values calculated from Equation (\ref{equation:2}) assuming $H = H_{scale}$ for all bodies, allowing for adjustments for superrotation with Venus and Jupiter. A least-squares regression through the origin yields a scaling $\phi_{scale}=b\phi_{obs}$ where $b$ has 95\% confidence bounds of $b=0.927\pm0.158$. The largest discrepancies between observed and predicted widths are observed with the gas giant planets, which have discrepancies ranging from -36\% to -64\% and from -8\degr~to -32\degr~latitude. The discrepancies are smallest for terrestrial bodies ranging from -1\% to -14\% and -1\degr~to -6\degr~(Table (\ref{table:2})). 
Assuming instead that $H = H_{trop}$ then  $b=1.21\pm0.285$. The largest discrepancies between observed and predicted widths are observed for the terrestrial planets, ranging from 11\% to 108\% and from 7\degr~to 65\degr~latitude. The discrepancies are smallest for gas giants ranging from -2\% to 38\% and -0.4\degr~to 11\degr~(Table (\ref{table:2})).

\begin{deluxetable*}{l|cc|cc|cc}[!htb]
\tablenum{2}
\tablecaption{Hadley Cell Width Calculated Using Equation (\ref{equation:2}) with $H_{scale}$ ($\phi_{scale}$), $H_{trop}$ ($\phi_{trop}$), and $H_{hf}$ ($\phi_{hf}$)}
\label{table:2}
\tablewidth{0pt}
\tablehead{
\colhead{Planet} & \colhead{$\phi_{scale}$} & \colhead{Discrepancy} & \colhead{$\phi_{trop}$} & \colhead{Discrepancy} & \colhead{$\phi_{hf}$} & \colhead{Discrepancy} \\  
\colhead{} & \colhead{(deg)} & \colhead{(\%)} & \colhead{(deg)} & \colhead{(\%)} & \colhead{(deg)} & \colhead{(\%)}
}
\startdata
Venus & 61.8 & 3 & 125.0 & 108 & 61.8 & 3\\
Earth & 26.0 & -13 & 36.8 & 23 & 26.0 & -13\\
Mars & 34.3 & -14 & 69.1 & 73 & 34.3 & -14\\
Jupiter & 10.1 & -43 & 22.0 & 22 & 22.1 & 23\\
Saturn & 11.5 & -54 & 24.6 & -2  & 23.5 & -6\\
Titan & 168.7 & -1 & 188.6 & 11  & 168.7 & -1\\
Uranus & 19.3 & -36 & 41.4 & 38 & 21.4 & -29\\
Neptune & 18.0 & -64 & 38.6 & -23 & 46.3 & -7\\
\enddata
\end{deluxetable*}

\section{Discussion} \label{sec:discussion}

There appears to be justification for applying Equation (\ref{equation:2}) beyond Earth to other planetary bodies in the solar system, although predicted cell widths can deviate significantly from observations, particularly when using $H_{scale}$ for gas giants and $H_{trop}$ for terrestrial bodies.

As a point of reference, we can define an effective atmospheric depth $H_{eff}$  that yields agreement between the calculated Hadley cell width using Equation (\ref{equation:2}) and the observed cell width, i.e., $H_{eff}=\Omega a \phi_H^{2}/N$. The relative value of $H_{eff}$ to $H_{trop}$ and $H_{scale}$ can then be quantified with a distance parameter:
\begin{equation}
d = \frac{H_{eff}-H_{scale}}{H_{trop}-H_{scale}}\label{equation:d}
\end{equation} 
shown in Table (\ref{table:3}) where a value of zero represents $H_{eff}$ = $H_{scale}$ and a value of 1 represents $H_{eff}$ = $H_{trop}$. Values greater than unity indicate an effective height greater than the tropopause height. 

\begin{figure*}[!htb]
\figurenum{2}
\plotone{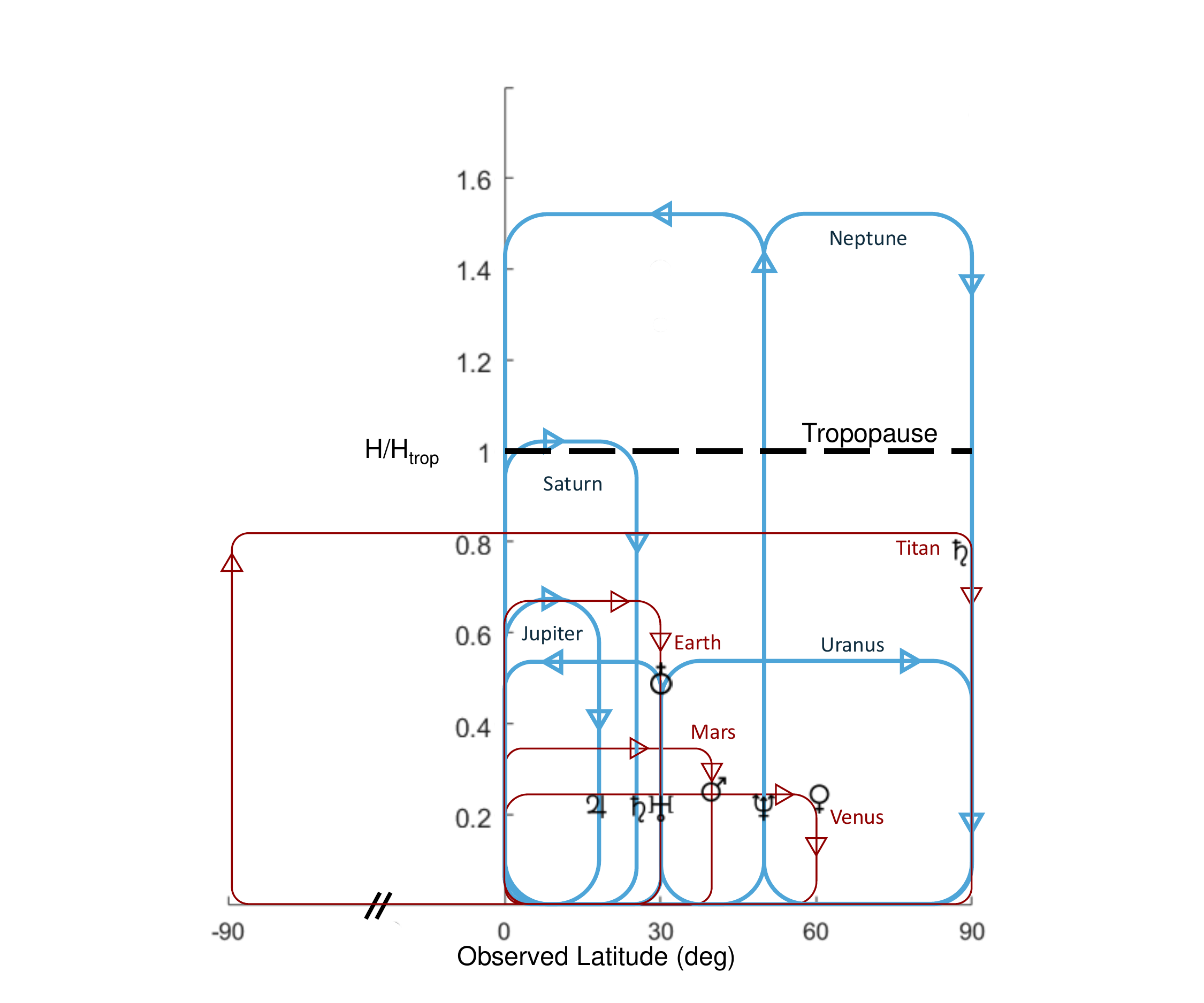}
\caption{Hadley cell circulation width from Equation (\ref{equation:2}) where $H_{eff}$ is scaled with respect to the tropopause (dashed black line) with $H_{scale}$ (planetary symbols) shown for comparison. Uranus and Neptune's dual cells are shown with sinking at both the poles and the equator. Titan's cell spans both hemispheres, with the Northern Hemisphere winter shown.}
\label{fig:fig2}
\end{figure*}

For terrestrial objects, $H_{eff}$ is closest to $H_{scale}$ with a mean $d$ value of 0.10 and a standard deviation of 0.15. With the exception of Uranus, gas giants have values of $H_{eff}$ closest to $H_{trop}$ with a mean value and standard deviation of $d=0.97\pm0.57$. Neptune has a value of $d$ considerably greater than unity, with a cell circulation that extends higher than the tropopause (Figure (\ref{fig:fig2})) consistent with recent modeling of Neptune's global circulation \citep{dePater}.  

\begin{figure}[!htb]
\figurenum{3}
\plotone{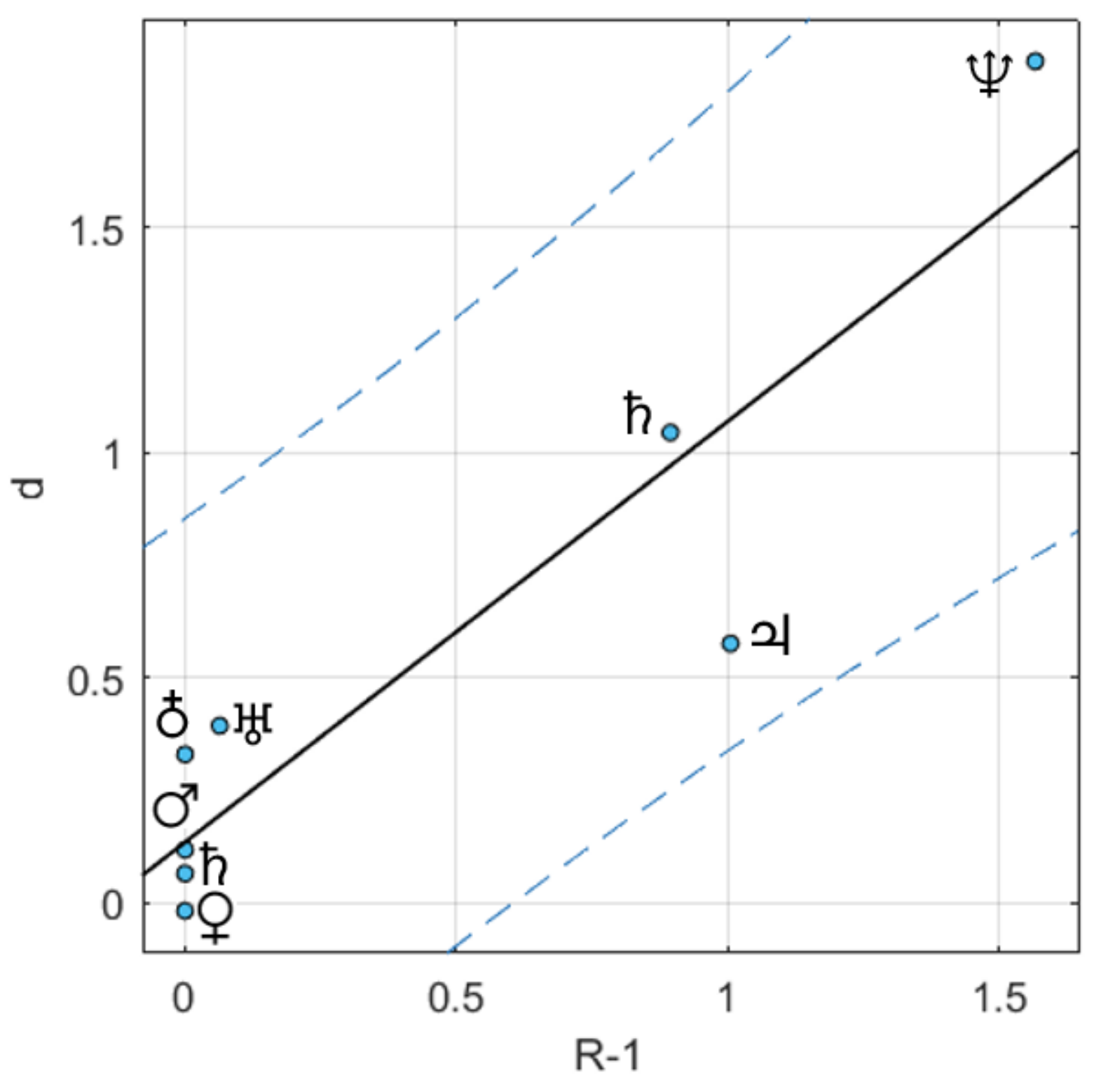}
\caption{A least-squares regression yields a best fit (solid line) with 95\% confidence bounds (dashed lines) of $d=b(R-1)+c$ where  $b=0.935\pm0.400$ and $c=0.134\pm0.292$. The Pearson correlation coefficient between $R-1$ and $d$ was found to be 0.92, where $p=0.001$}
\label{fig:dvrm1}
\end{figure}

Why is it that gas giants generally have higher values of $d$? One notable distinction is that they are characterized by an internal heat source arising from Kelvin-Helmholtz contraction due to compression and interior heating \citep{Guillot}.  Expressing the planetary absorbed shortwave flux as $F_{sw}=(1-\alpha)S_{0}/4$ where $\alpha$ is the bond albedo \citep{NASA,Li2011,Li2018} and $S_{0}$ is the solar flux at the top of the atmosphere \citep{NASA,Li2011,Li2018}, and the outgoing longwave flux as $F_{lw}=\sigma T_{e}^{4}$ where $\sigma$ is the Stefan-Boltzmann constant and $T_{e}$ is the global blackbody emission temperature \citep{Showman}, then the ratio of the emitted longwave heat flux to the absorbed solar flux is 
\begin{equation}
R = F_{lw}/F_{sw} \label{fluxratio} 
\end{equation}
Global values are used for the latent heat flux and outgoing longwave radiation so that $R$ is independent of the characteristics of the Hadley cell. For example, on Earth, there is an imbalance in the tropics between incoming and outgoing radiation; equilibrium temperatures are maintained due to the meridional heat flux out of the tropics in large part due to the Hadley Cell. 

\begin{figure*}[!htb]
\figurenum{4}
\centering
\plotone{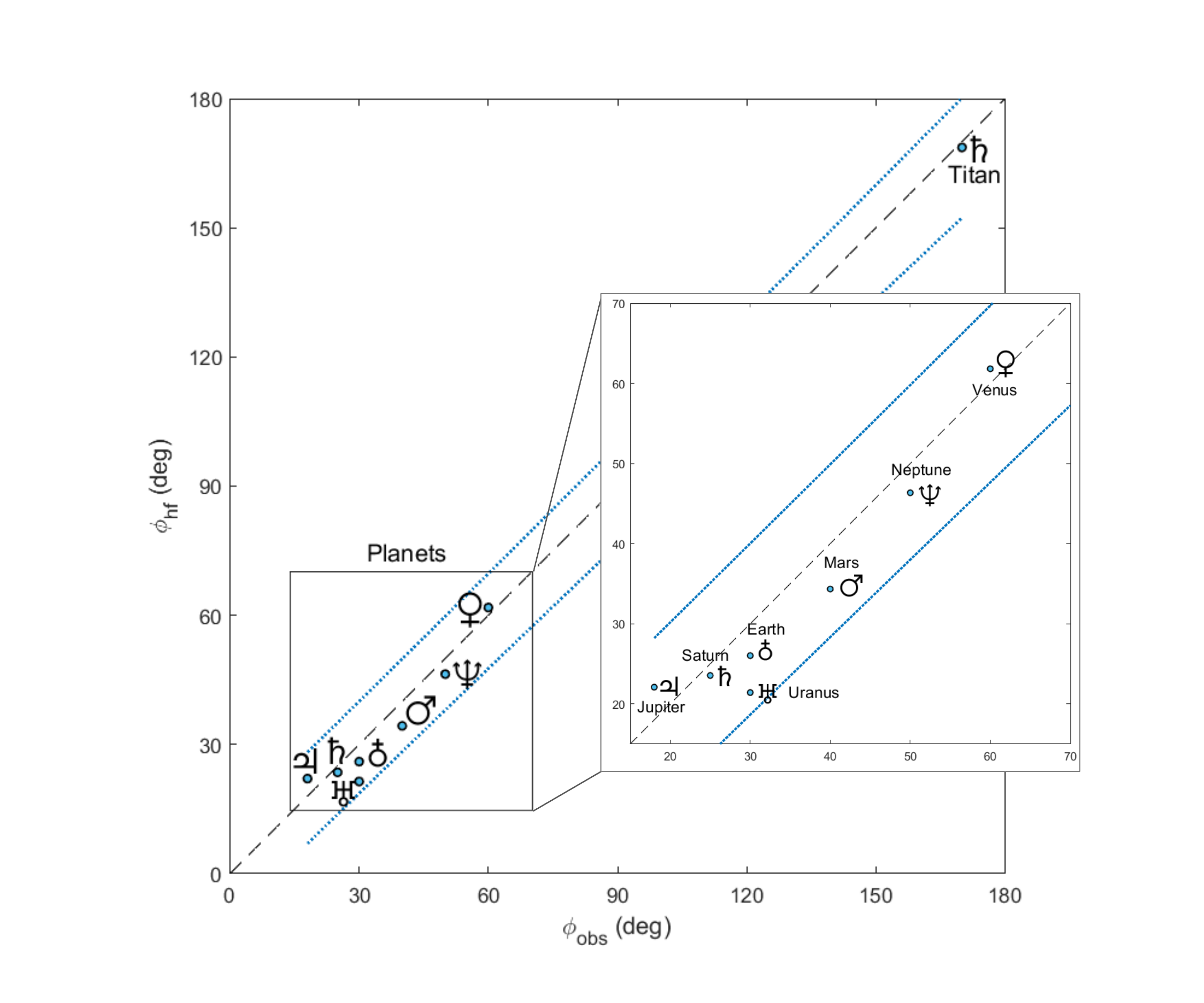}
\caption{Observed values of Hadley cell width versus values calculated with Equation (\ref{equation:2}) using a value of $H$ adjusted for the heat flux ratio according to Equation (\ref{equation:Hhf}). Blue dotted lines show 95\% confidence interval bounds for a least-squares best fit of $\phi_{hf}=b\phi_{obs}$ where $b=0.978\pm0.084$. Dashed lines are where $\phi_{obs}=\phi_{hf}$.}
\label{fig:fig3}
\end{figure*}

Each of the terrestrial bodies can be assumed to be in radiative equilibrium with no significant internal heat source and $R=1$ \citep{Li2011, Avduevsky}. For the gas giants, as shown in Table (\ref{table:3}), excess internal heat is largest on Neptune with $R=2.57$ while Uranus is more similar to terrestrial planets with $R=1.06$. Figure (\ref{fig:dvrm1}) shows a high degree of linear correlation between $d$ and $R-1$ suggesting an adjustment to the circulation height that accounts for the internal heat flux given by
\begin{equation}
H_{hf} = H_{scale} + (R-1)(H_{trop} - H_{scale}) \label{equation:Hhf}
\end{equation}
As shown in Table (\ref{table:2}), using $H_{hf}$ in place of $H_{scale}$ appears to reduce discrepancies between calculated and observed values for the Hadley cell width. The largest discrepancies by percentage from this method are Uranus (-29\%) and Jupiter (23\%) but all planets' calculated cell widths are within 10\degr~latitude of their observed values. As shown in Figure (\ref{fig:fig3}), the revised coefficient $b$ relating calculated and observed values $\phi=b\phi_{obs}$ is $b=0.978\pm0.084$.

While Earth is a terrestrial planet with $R = 1$, it nonetheless has a value of $H_{eff}$ that is significantly higher than $H_{scale}$ with a value of $d=0.33$, relatively large compared to  other terrestrial planets. Earth is not characterized by an internal heat source from Kelvin-Helmholtz contraction as with the gas giants. However, deep convection in the tropics is driven by an unusually high degree of latent heat release \citep{Read, Williams}. Recent refinements to Earth's global energy budget suggest that the latent heat flux is $88\pm10$ W m$^{-2}$ or approximately one third of the outgoing longwave flux \citep{Stephens}. By comparison, latent heat release on Titan from the methane cycle constitutes on the order of 0.01\% of $F_{lw}$ \citep{Williams}, and on Mars latent heat release from the CO2 cycle is only about 1\% of the total energy budget \citep{Read}.

If latent heat release $LH$ is taken to act as an internal heat source that pushes the upper boundary of Earth's Hadley circulation upwards, then $R-1=LH/F_{lw} = 0.37$, closely corresponding to the inferred value of $d=0.33$. 

\begin{deluxetable*}{lccccccccc}[!htb]
\tablenum{3}
\tablecaption{Hadley Cell Effective Height $H_{eff}$, Physical Constants for Heat Source Determination, and Distance Parameter $d$ (Equation (\ref{equation:d}))}
\label{table:3}
\tablehead{
\colhead{Planet} & \colhead{$H_{eff}$} & \colhead{$H_{hf}$} & \colhead{$\alpha$} & \colhead{$S_{0}$} & \colhead{$T_{e}$} & \colhead{$F_{lw}$} &\colhead{$F_{sw}$} & \colhead{$R$} & \colhead{$d$}\\ 
\colhead{} & \colhead{(km)} & \colhead{(km)} & \colhead{} & \colhead{(W~m$^{-2}$)} & \colhead{(K)} & \colhead{(W~m$^{-2}$)} & \colhead{(W~m$^{-2}$)} & \colhead{} & \colhead{}
}
\startdata
Venus & 15.0 & 15.9 & 0.75 & 2601.3 & 232 & 164.26 & 162.58 & 1.00 & -0.02 \\
Earth & 11.3 & 8.50 & 0.31 & 1361.0 & 255 & 239.74 & 238.18 & 1.00 & 0.33  \\
Mars & 15.1 & 11.1 & 0.25 & 586.2 & 210 & 110.27 & 109.91 & 1.00 & 0.12  \\
Jupiter & 83.0 & 125 & 0.50 & 53.5 & 124 & 13.41 & 6.69 & 2.01 & 0.57  \\
Saturn & 283 & 251 & 0.34 & 14.8 & 95 & 4.62 & 2.44 & 1.89 & 1.04  \\
Titan & 40.6 & 40.0 & 0.26 & 15.2 & 85 & 2.96 & 2.80 & 1.00 & 0.06 \\
Uranus & 66.9 & 34.1 & 0.30 & 3.7 & 59 & 0.69 & 0.65 & 1.06 & 0.39  \\
Neptune & 155 & 133 & 0.29 & 1.5 & 59 & 0.69 & 0.27 & 2.57 & 1.87  \\
\enddata
\end{deluxetable*}

The implication is that internal heat sources and latent heat influence the Hadley cell width through convective processes and cause the atmospheric circulation to deviate upwards from the pressure scale height.  It is also possible that some remaining measure of any observed discrepancies is due simply to uncertainties in observations of  $\phi_{obs}$ and $N$. Tropospheric temperature profiles for Neptune and Uranus exist only from radio occultation data \citep{Lunine}, and the global circulation pattern for Neptune has been inferred only indirectly by telescopes at the infrared and radio wavelengths \citep{dePater}. Perhaps revealingly, it was not until the release of recent \textit{Juno} and \textit{Cassini} probe data for Jupiter and Saturn that the existence of deeply penetrating atmospheric jets became known \citep{Kaspi} as well as the latitudinal locations of atmospheric downwelling \citep{Friedson, Bolton}. Using older data for Saturn's observed Hadley cell width and superrotation on Jupiter, the difference between theory and observations for Saturn and Jupiter is -26\% and 23.2\%, respectively, compared with -6\% and 22.7\% using the newer data sets.

\section{Conclusions} \label{sec:conclusions}
An analytical expression for Hadley cell latitudinal width on Earth given by Equation (\ref{equation:2}) appears to provide estimates for other solar system planetary bodies that agree well with observations provided that an account is made where necessary for atmospheric super-rotation and an adjustment is made to the pressure scale height for any internal heat source. As measurements of Earth-like exoplanets improve \citep{KaspiShowman}, Equation (\ref{equation:2}) may provide guidance as to their general circulation patterns and possible habitability.


\clearpage

\bibliographystyle{aasjournal}
\bibliography{planetsref}

\end{document}